%% file: main.tex
\documentclass[acmtog,nonacm]{acmart}
\usepackage{colortbl}
\usepackage{multirow}
\usepackage{varwidth} 
\usepackage{pifont} 
\usepackage{boldline}
\usepackage{amsmath}

\usepackage{xcolor}
\usepackage[ruled]{algorithm2e} 
\usepackage{wrapfig}
\usepackage{bm}
\usepackage{makecell}
\usepackage{arydshln}
\usepackage{siunitx}
\usepackage{afterpage}
\usepackage{graphicx}  
\usepackage{float}     
\usepackage[shortcuts]{extdash}
\usepackage[utf8]{inputenc}
\usepackage{geometry}
\usepackage{booktabs} 
\usepackage{array}
\usepackage{caption}
\usepackage{makecell}
\usepackage{xcolor}
\usepackage{amssymb} 
\usepackage{pifont}  

\usepackage{tikz}        
\usepackage{xcolor}      
\usepackage{pifont}      
\usepackage{amssymb}     
\usepackage{array}       

\newcolumntype{P}[1]{>{\centering\arraybackslash}p{#1}}
\AtBeginDocument{%
  }

\setcopyright{acmlicensed}
\copyrightyear{2025}
\acmYear{2025}
\acmDOI{XXXXXXX.XXXXXXX}

\acmJournal{TOG}
\acmVolume{44}
\acmNumber{4}
\acmArticle{111}
\acmMonth{8}

\acmSubmissionID{1221}

\begin{document}
\input{definitions}

\title{Fast Wave-optics Rendering of Multiplane Images for 3D Holographic Displays}

\author{Brian Chao}
\authornote{Corresponding author.}
\email{brianchc@stanford.edu}
\affiliation{%
  \institution{Stanford University}
  \country{USA}
}
\affiliation{%
  \institution{Meta}
  \country{USA}
}

\author{Dario Seyb}
\email{dseyb@meta.com}
\affiliation{%
  \institution{Meta}
  \country{USA}
}

\author{Nathan Matsuda}
\email{nathan.matsuda@meta.com}
\affiliation{%
  \institution{Meta}
  \country{USA}
}

\author{Oliver Cossairt}
\email{ocossairt@meta.com}
\affiliation{%
  \institution{Meta}
  \country{USA}
}

\author{Yang Zhou}
\email{zhouyang@meta.com}
\affiliation{%
  \institution{Meta}
  \country{USA}
}

\author{Douglas Lanman}
\email{douglas.lanman@meta.com}
\affiliation{%
  \institution{Meta}
  \country{USA}
}

\author{Gordon Wetzstein}
\email{gordon.wetzstein@stanford.edu}
\affiliation{%
  \institution{Stanford University}
  \country{USA}
}

\author{Grace Kuo}
\email{gracekuo@meta.com}
\affiliation{%
  \institution{Meta}
  \country{USA}
}

\author{Changwon Jang}
\email{changwon.jang@meta.com}
\affiliation{%
  \institution{Meta}
  \country{USA}
}

\renewcommand{\shortauthors}{Chao et al.}

\begin{abstract}
%

Recent advances in neural rendering have unlocked unprecedented capabilities in 3D reconstruction and novel view synthesis, giving rise to applications such as virtual fly-throughs of a 3D scene reconstructed from a set of sparse, casually captured images. However, these renderings are viewed on a computer screen or conventional VR headsets as 2D images, greatly limiting the perceptual realism and immersiveness of such experiences. The rapid development in novel 3D scene representations calls for dedicated rendering algorithms that convert these readily-available 3D contents into formats that are compatible with emerging 3D display technologies, such as holographic displays. In this paper, we propose a wave-optics rendering pipeline that works with multiplane images (MPIs) for efficient and high-quality hologram synthesis. Our MPI-based computer-generated holography algorithm greatly outperforms state-of-the-art primitive-based CGH algorithms in terms of runtime, achieving speedups up to $250,000\times$ while achieving comparable image quality, and significantly outperforms conventional layer-based CGH algorithms in terms of image quality. We validate our method extensively on a wide variety of 3D scene datasets both in simulation and through experimentally captured results, showing exceptional 3D focal stack and 4D light field reconstruction performance without sacrificing efficiency.
\end{abstract}




\received{23 May 2025}
\received[revised]{10 August 2025}
\received[accepted]{10 August 2025}

\begin{teaserfigure}
  \centering
	\includegraphics[width=\columnwidth]{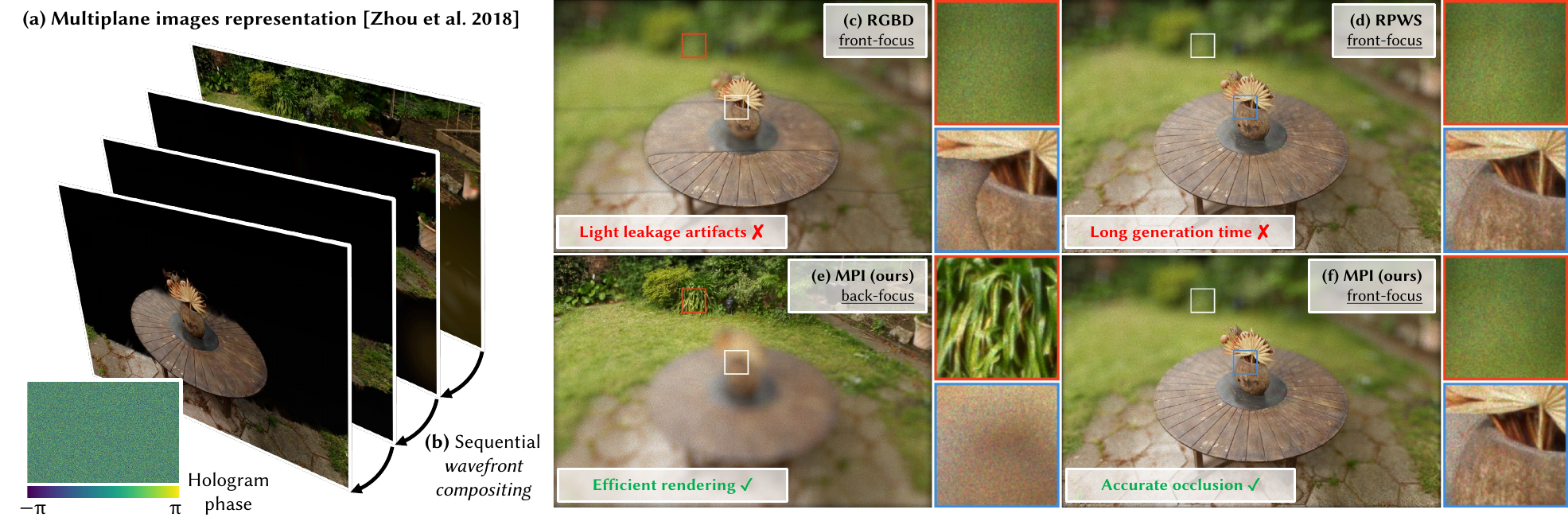}
   \caption{We propose a wave-optics rendering pipeline for multiplane images (MPIs) \textbf{(a, b)} to generate holograms that accurately reconstruct 3D focal stacks and 4D light fields. Conventional RGBD holograms are fast to render, yet suffer from incorrect defocus and parallax behaviour at depth transitions \textbf{(c)}. The state-of-the-art primitive-based CGH algorithm, RPWS \cite{chao2025rpws}, generates high-quality holograms but suffer from extremely slow runtime \textbf{(d)}. Our proposed MPI-based CGH efficiently converts MPIs into holograms that accurately reconstruct 3D focal stacks and 4D light fields \textbf{(e), (f)}. The randomness of the final composited hologram, as shown in the color-coded phase map on the left, significantly expands the eyebox and improves the space--bandwidth product utilization of the spatial light modulator in the holographic display, allowing for natural defocus and accurate parallax reconstruction.}
  \label{fig:teaser}
\end{teaserfigure} 

\maketitle

\section{Introduction}
\input{main_paper/sections/1_intro}

\section{Related Work}
\input{main_paper/sections/2_previous_work}

\section{MPI-based Computer-generated Holography}
\input{main_paper/sections/3_method}

\section{Experiments and Analysis}
\label{sec:results}
\input{main_paper/sections/4_results}

\section{Discussion}
\label{sec:discussion}
\input{main_paper/sections/5_discussion}

\bibliographystyle{ACM-Reference-Format}
\bibliography{bibs}


\begin{figure*}[ht!]
    \centering
    \includegraphics[width=0.95\textwidth]{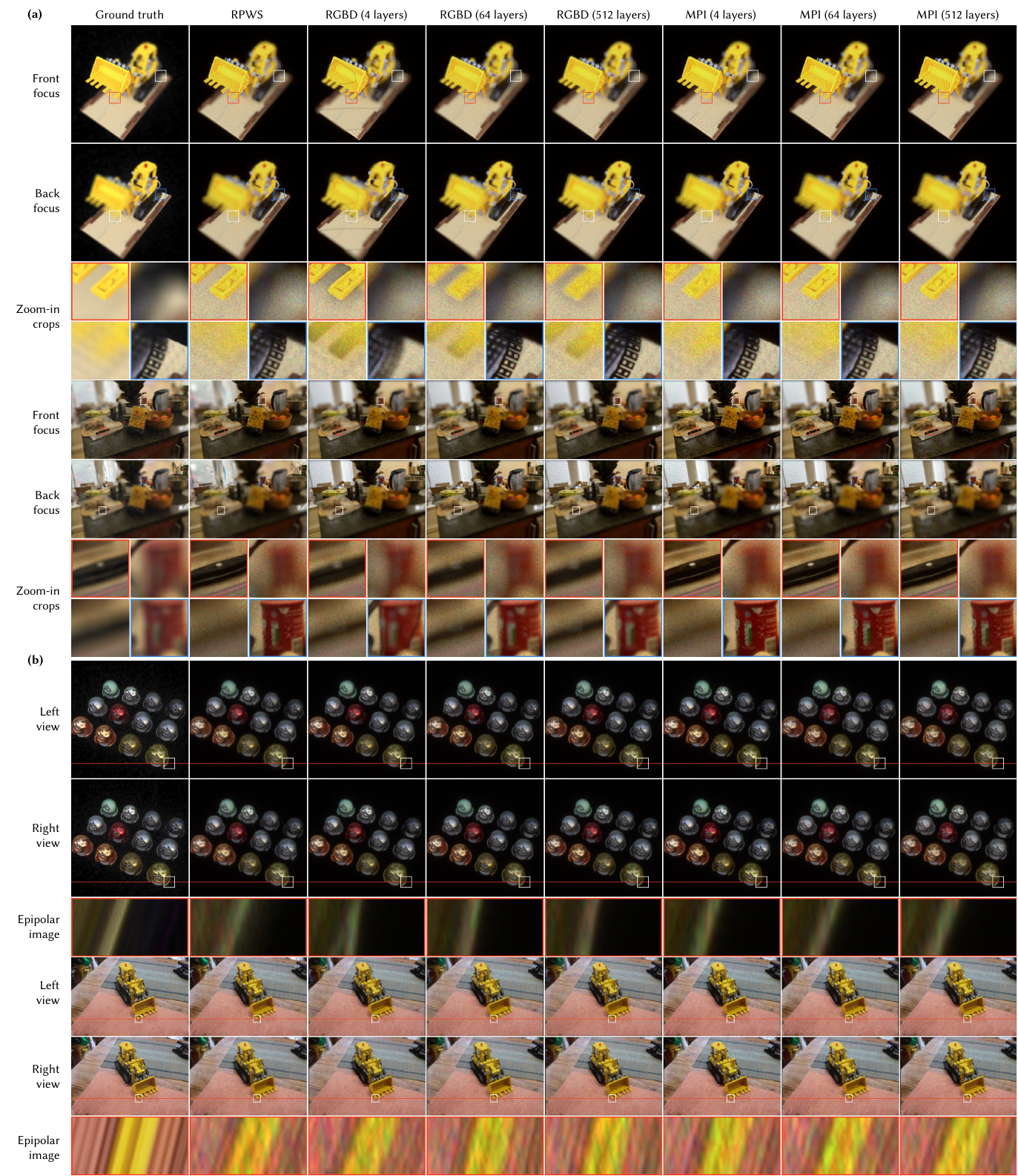}
    \caption{\textbf{Simulated 3D focal stacks and 4D light fields of holograms generated using different CGH algorithms. } RPWS achieves state-the-of-art focal stack and light field reconstruction quality among primitive-based CGH methods. RGBD holograms, despite being efficient to render, suffer from light leakage and halo artifacts in the reconstructed focal stacks and incorrect occlusion in the reconstructed light fields at depth transitions. Increasing the depth layers slightly improves visual quality, but the artifacts cannot be fully eliminated. Our MPI holograms achieve image quality comparable to RPWS, while offering up to $250,000\times$ faster rendering speed (for 4 MPI layers), with the exact speedup being proportional to the number of MPI layers. }
    \label{fig:results_sim}
\end{figure*}

\begin{figure*}[ht!]
    \centering
    \includegraphics[width=0.95\textwidth]{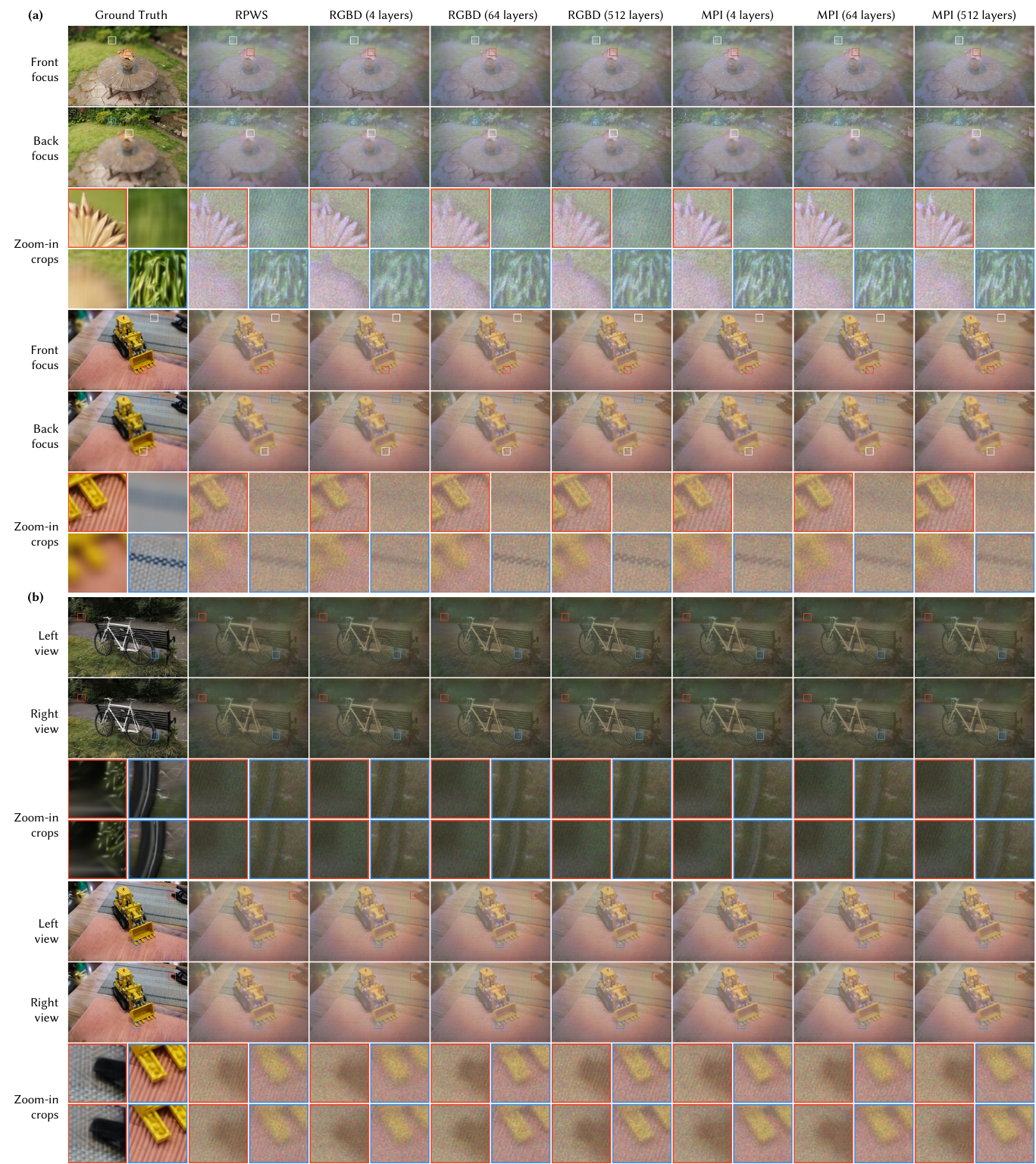}
    \caption{\textbf{Experimentally captured 3D focal stacks and 4D light fields of holograms generated using different CGH algorithms. } We capture 3D focal stacks and horizontal parallax results of holograms generated using different CGH algorithms on our benchtop holographic display setup. The trend in the captured results well match the simulated experiments. Both RPWS and our MPI holograms reconstruct natural defocus blur and parallax. RGBD holograms suffer from artifacts at depth transitions, as shown in the garden table zoom-in crops of RGBD holograms in \textbf{(a)}. }
    \label{fig:results_captured}
\end{figure*}

\end{document}

%% file: definitions.tex
\newcommand{\rref}{\mathbf{r}_{\text{0}}}
\newcommand{\xref}{{x}_{\text{0}}}
\newcommand{\yref}{{y}_{\text{0}}}
\newcommand{\zref}{{z}_{\text{0}}}

\newcommand{\rlocal}{\mathbf{r}_{l}}
\newcommand{\xlocal}{{x}_{l}}
\newcommand{\ylocal}{{y}_{l}}
\newcommand{\zlocal}{{z}_{l}}

\newcommand{\cov}{\mathbf{\Sigma}}
\newcommand{\rot}{\mathbf{R}}
\newcommand{\rotc}{\mathbf{C}}
\newcommand{\sca}{\mathbf{S}}
\newcommand{\kvec}{\mathbf{k}}

\definecolor{mg}{rgb}{0.639,0.984,0.722}
\definecolor{my}{rgb}{0.996,0.875,0.643}
\definecolor{mr}{rgb}{0.941,0.561,0.620}
\newcommand{\ccg}{\cellcolor{mg}}  
\newcommand{\ccy}{\cellcolor{my}}  
\newcommand{\ccr}{\cellcolor{mr}}  

\newcommand{\cmark}{\ding{51}}%
\newcommand{\xmark}{\ding{55}}%
\newcommand{\bluecheck}{{\color{blue}\checkmark}}
\newcommand{\blackcheck}{{\color{black}\cmark}}
\newcommand{\redx}{{\Large\color{\red}\xmark} }

\newcommand{\headfontsize}{\scriptsize}
\newcommand{\tablerefsize}{\tiny}
\newcommand{\vertice}{\mathbf{v}}
\newcommand{\normal}{\mathbf{n}}

\newcommand{\roundedtriangle}{%
  \begin{tikzpicture}[baseline=0ex, scale=0.06]
    \draw[thick] (1.5,1.5) circle (1.5);
  \end{tikzpicture}%
}
\newcommand*\colourcheck[1]{%
  \expandafter\newcommand\csname #1check\endcsname{\textcolor{#1}{\ding{51}}}%
}
\colourcheck{red}
\colourcheck{green}

\newcommand*\colourxx[1]{%
  \expandafter\newcommand\csname #1xx\endcsname{\textcolor{#1}{\ding{55}}}%
}
\colourxx{red}
\colourxx{green}

\newcommand*\colourtriangle[1]{%
  \expandafter\newcommand\csname #1triangle\endcsname{\textcolor{#1}{$\varDelta$}}%
}
\colourtriangle{yellow}


\definecolor{statusgreen}{RGB}{34, 139, 34}   
\definecolor{statusorange}{RGB}{218, 165, 32} 
\definecolor{statusred}{RGB}{178, 34, 34}     
\definecolor{rowhighlight}{RGB}{240, 248, 255} 

\newcommand{\resGood}{\textcolor{statusgreen}{\ding{52}}}       
\newcommand{\resAvg}{\textcolor{statusorange}{\textbf{\roundedtriangle}}} 
\newcommand{\resBad}{\textcolor{statusred}{\ding{56}}}          

\newcommand{\greenvcell}{\resGood}
\newcommand{\yellowtrcell}{\resAvg}
\newcommand{\redxcell}{\resBad}


\newcommand{\ctcell}{>{\centering\arraybackslash}}
\newcommand{\gw}[1]{\textcolor{purple}{[GW: #1]}}
\newcommand{\sy}[1]{\textcolor{blue}{[SC: #1]}}
\newcommand{\bc}[1]{\textcolor{red}{[BC: #1]}}
\newcommand{\jy}[1]{\textcolor{olive}{[JY: #1]}}
\newcommand{\edit}[1]{\textcolor{red}{#1}}


\newcommand{\citl}{CITL}
\newcommand{\revised}[1]{{#1}}
\newcommand{\editsy}[1]{\textcolor{red}{#1}}
\newcommand{\prop}{\mathcal{P}}
\newcommand{\firstprop}{\prop_1}
\newcommand{\secprop}{\prop_2}
\newcommand{\prophat}{\widehat{g}}
\newcommand{\firstprophat}{\prophat_1}
\newcommand{\secprophat}{\prophat_2}
\newcommand{\target}{a_{target}}
\newcommand{\phaseslm}{\phi}
\newcommand{\efficiency}{\eta}
\newcommand{\fourier}{\mathcal{F}}
\newcommand{\loss}{\mathcal{L}}
\newcommand{\transfer}{\mathcal{H}}
\newcommand{\fx}{f_x}
\newcommand{\fy}{f_y}
\newcommand{\numpixelx}{N_x}
\newcommand{\numpixely}{N_y}
\newcommand{\phasepool}{\mathbb{P}}
\newcommand{\capturedpool}{\mathbb{C}}
\newcommand{\rgbd}{\mathbb{RGBD}}
\newcommand{\lossbp}{\texttt{loss\_bp}}
\newcommand{\fbp}{\texttt{f\_bp}}
\newcommand{\gbp}{\texttt{g\_bp}}
\newcommand{\red}[1]{\textcolor{red}{#1}}
\newcommand{\idx}{i}
\newcommand{\maxidx}{N}
\newcommand{\angularspectrum}{\hat{u}}
\newcommand{\pixelcoord}{\mathbf{r}}
\newcommand{\transmittance}{\mathcal{T}}

\newcommand{\coordinate}{\mathbf{x}}
\newcommand{\freqcoords}{\mathbf{k}}
\newcommand{\mean}{\bm{\mu}}

\newcommand{\worldspace}{w}
\newcommand{\viewspace}{r}
\newcommand{\canonicalspace}{c}
\newcommand{\objectspace}{o}
\newcommand{\centergaussian}{{\bm{\mu}}}
\newcommand{\projectivemapping}{\mathbf{m}}

\newcommand{\mat}[1]{\mathbb{R}^{{#1}\times{#1}}}
\newcommand{\arr}[1]{\mathbb{R}^{{#1} \times 1}} 

\makeatletter
\newcommand\xleftrightarrow[2][]{%
  \ext@arrow 9999{\longleftrightarrowfill@}{#1}{#2}}
\newcommand\longleftrightarrowfill@{%
  \arrowfill@\leftarrow\relbar\rightarrow}
\makeatother

%% file: main_paper/sections/1_intro.tex
Holographic near-eye displays are an emerging class of 3D display technology that enable realistic, immersive visual experiences through physically accurate wavefront reconstruction \cite{maimone2017holographic, chao2024large, gopakumar2024full, kim:HolographicGlasses, jang2024waveguide}. However, despite their immense promise, their practical deployment remains fundamentally constrained by the extreme computational cost of computer-generated holography (CGH). Driving such displays requires synthesizing interference patterns from complex 3D inputs in real time, yet this mapping from scene representation to SLM wavefront modulation is computationally demanding.

Existing CGH algorithms work with a wide range of 3D representations. However, each representation introduces significant computational trade-offs that limit real-time CGH:
RGBD images \cite{choi2021neural, peng2020neural} suffer from occlusion artifacts, mesh- \cite{matsushima2005computer, matsushima2009extremely}, point-cloud- \cite{h2009computer, maimone2017holographic}, or Gaussian-based \cite{choi2025gaussian, chao2025rpws} pipelines must process millions of primitives, and densely-sampled light fields \cite{chao2024large, schiffers2023stochastic} require content that is rarely produced in typical capture settings. As a result, generating high-quality, photorealistic holograms at interactive rates from compact or casually captured inputs remains an open challenge. 

Multiplane images (MPIs) \cite{zhou2018mpi, pu2023sinmpi, khan2023tiledmpi, han2022adaptivempi, tucker2020singlempi, flynn2019deepviewmpi, zhao2022generativempi, luvizon2021adaptivempi, srinivasan2019pushingmpi} are a layered 3D representation that consists of a set of RGBA images, each defined at a corresponding depth which can be predetermined or adaptively optimized from a set of casually captured images of a front-facing 3D scene. Novel views are synthesized through warping the RGBA images from a reference viewpoint to the novel viewpoint and subsequent alpha compositing. Although a single set of MPIs is unable to achieve 360-degree, free-view 3D reconstruction due to its fronto-parallel arrangement, it provides exceptional reconstruction of novel views particularly for \textit{local, front-facing light fields} of a 3D scene. This property is especially well-suited for CGH since only front-facing views need to be reconstructed well in the small eyebox of a near-eye display. Most importantly, MPIs usually only consist of a handful of RGBA images (<100), allowing for much faster processing in the wave optics domain compared to Gaussians and meshes where millions of primitive wavefronts have to be propagated and composited \cite{choi2025gaussian, matsushima2009extremely, matsushima2018full}.

In this paper, we devise a CGH pipeline that uses MPIs as an intermediate 3D representation for efficient computer-generated holography. We show that MPI-based CGH achieves high-quality focal stack and light field reconstruction quality and outperforms other conventional layer-based CGH methods such as CGH using RGBD images. We further demonstrate that MPI-based CGH can achieve up to $250,000\times$ speedup compared to state-of-the-art primitive-based methods such as polygon- or Gaussians-based CGH \cite{matsushima2005computer, matsushima2009extremely, matsushima2014silhouette, choi2025gaussian}, paving the way for practical and real-time CGH.

Specifically, our contributions include:
\begin{itemize}
\item A novel CGH algorithm that composites multiplane images into complex holograms with optional time multiplexing.
\item A detailed evaluation of MPI-based holograms against existing CGH baselines, analyzing how the number of depth layers impacts image quality and runtime.
\item Experimental validation demonstrating accurate 3D focal stack reconstruction and \textit{4D light field parallax} in MPI-based holograms.
\end{itemize}


Source code and example datasets will be made public.

%% file: main_paper/sections/2_previous_work.tex
CGH has a long history of algorithmic and display innovations. In the next section, we highlight prior work most relevant to our approach and point interested readers to broader surveys of holographic displays~\cite{yaracs2010state,park2017recent,chang2020toward,javidi2021roadmap,pi2022review}.

\paragraph{\bf{Computer-generated Holography}}

Computer-generated holography (CGH) is a process that maps a three-dimensional volume onto the 2D plane of a spatial light modulator (SLM) via simulated interference. This computational process is heavily dependent on the format of the input data; historically, CGH frameworks have been designed to handle inputs ranging from explicit geometric primitives, such as point clouds, meshes, and Gaussians, to layered representations like RGBD images and densely-sampled image sets like light fields and focal stacks.

\renewcommand{\arraystretch}{1.1}
\setlength{\tabcolsep}{4pt} 
\setlength{\textfloatsep}{10pt}

\begin{table}[t!]
    \centering
    \captionsetup{aboveskip=0pt, belowskip=3pt}
    {\footnotesize
    \caption{\textbf{Comparison of different 3D representations for CGH in terms of efficiency and visual quality.} Point cloud and mesh representations lead to poor appearance reconstruction quality due to the underlying coarse 3D representation. CGH algorithms based on focal stack and light field targets require iterative optimization. RGBD holograms are efficient to synthesize, yet image quality degrades at depth transitions. Gaussian-based holograms, on the other hand, achieve superior quality but suffers from slow generation time. Our MPI-based holograms achieve the best of both worlds of RGBD and Gaussian holograms, simultaneously achieving efficient runtime and high image quality. }
    \label{tab:cgh_comparison}

    \begin{tabular}{
        >{\raggedright\arraybackslash}m{2.8cm} 
        >{\centering\arraybackslash}m{1.5cm} 
        >{\centering\arraybackslash}m{1.5cm} 
        >{\centering\arraybackslash}m{1.5cm}
    }
        \toprule 
        \textbf{Methods} & 
        \makecell{\textbf{Efficient}\\\textbf{inference}} & 
        \makecell{\textbf{Focal stack}\\\textbf{quality}} & 
        \makecell{\textbf{Light field}\\\textbf{quality}} \\ 
        \midrule 
        
        Point cloud & \yellowtrcell & \redxcell & \redxcell \\
        \addlinespace 
        
        RGBD & \greenvcell & \yellowtrcell & \yellowtrcell \\
        \addlinespace
        
        Focal stack & \yellowtrcell & \greenvcell & \yellowtrcell \\
        \addlinespace
        
        Light field & \yellowtrcell & \greenvcell & \greenvcell \\
        \addlinespace
        
        Mesh & \redxcell & \redxcell & \redxcell \\
        \addlinespace
        
        Gaussians  & \redxcell & \greenvcell & \greenvcell \\
        
        \midrule 
        \rowcolor{rowhighlight} 
        \textbf{MPIs} & \greenvcell & \greenvcell & \greenvcell \\
        \bottomrule 
    \end{tabular}
    }
\end{table}

Regardless of the input representation, the computational efficiency of CGH algorithms is fundamentally determined by the choice between optimization-based and direct algorithms. Optimization-based approaches, while capable of reconstructing focal stacks and light fields with high fidelity through error minimization ~\cite{chakravarthula2019wirtinger, zhang20173d, peng2020neural, fienup1980iterative, gerchberg1972practical, schiffers2023stochastic}, suffer from prohibitive runtime costs. These methods require tens to hundreds of wave propagation steps and gradient updates per frame, making them unsuitable for real-time operation.

In contrast, direct methods offer a pathway to interactive frame rates by synthesizing holograms in a single forward computation. However, the efficiency of direct methods is strictly bounded by the complexity of the input representation. Methods that model scenes with primitives, such as points, triangular meshes, or Gaussians, must propagate millions of individual wavefronts and employ complex compositing schemes for occlusion~\cite{matsushima2014silhouette, choi2025gaussian, chao2025rpws}, often reintroducing significant latency. To address this, learning-based variants use neural networks to predict holograms directly from 3D inputs~\cite{shi2021towards, yang2022diffraction}, though often at the cost of high memory usage or training overhead. This highlights the need for direct algorithms that operate on structurally efficient representations to maximize throughput. Please refer to Table \ref{tab:cgh_comparison} for a detailed comparison of CGH algorithms based on different 3D representations.

In this work, we develop a \textit{direct} CGH pipeline that transforms multiplane images (MPIs) 3D representations into complex holograms, offering a practical balance between quality and efficiency. We distinguish this algorithm from \textit{phase- or amplitude-encoding} techniques that map complex holograms to phase-only or amplitude-only patterns for standard SLMs, including double-phase amplitude coding (DPAC) \cite{dpac}, single-sideband encoding \cite{sideband}, stochastic optimization \cite{peng2020neural}, and learned phase-retrieval networks \cite{shi2021towards}.

\paragraph{\bf{3D Scene Representations}} The selection of a 3D scene representation for CGH is governed by an explicit trade-off between reconstruction fidelity and computational efficiency. Simple representations favor efficiency but compromise on quality. For example, point clouds~\cite{maimone2017holographic} are fast to process but inherently sparse, resulting in holes in reconstruction. On the other hand, RGBD images~\cite{choi2021neural, peng2020neural} enable rapid hologram synthesis but lack continuous volumetric information, leading to severe artifacts at occlusion boundaries.

Other geometric primitive representations prioritize visual quality at the expense of runtime. While Gaussian splats~\cite{choi2025gaussian, chao2025rpws} provide more geometric accuracy and photorealism, they incur massive computational costs due to the need to process millions of individual primitives and handle their occlusions. Other formats, such as densely-sampled light fields~\cite{schiffers2023stochastic, choi2022time, kim2024holographic}, offer high fidelity but face fundamental spatio–angular resolution tradeoffs and immense data requirements. Consequently, identifying a representation that bridges the gap between the efficiency of sparse and layered inputs and the visual quality of densely populated primitives remains an open challenge.

To address this trade-off between efficiency and quality, we propose leveraging \textit{multiplane images (MPIs)} as an effective 3D representation for CGH. MPIs combine the efficiency of layered RGBD-style depth rendering with the continuous volumetric expressiveness of Gaussian representations. Each slice of an MPI is a plane with RGBA information. Novel views can be rendered by applying planar homographies to each layer and compositing the warped RGBA images using alpha blending~\cite{zhou2018mpi}.

MPIs have been shown to reconstruct front-facing scenes and moderate parallax with high fidelity using only a small number of layers (typically fewer than 100), and can be synthesized in real time with feed-forward neural networks~\cite{pu2023sinmpi, khan2023tiledmpi, han2022adaptivempi, tucker2020singlempi, flynn2019deepviewmpi, zhao2022generativempi, luvizon2021adaptivempi, srinivasan2019pushingmpi, zhou2018mpi}. This characteristic is particularly well suited for holographic near-eye displays, which only require accurate reconstruction of front-facing views over a limited eyebox size rather than full 360° scene coverage as in Gaussian splatting. These properties position MPIs as an ideal intermediate representation for computer-generated holography. We demonstrate that MPIs achieve the best of both worlds, matching the visual fidelity of Gaussian-based holograms while maintaining the efficiency of RGBD-based approaches.

\paragraph{\bf{Layer-based 3D Displays and Scene Representations}} There has been extensive prior research on \textit{layer-based, glasses-free 3D displays} that aim to reconstruct light fields directly to the user's eyes~\cite{lanman2010hr3d, wetzstein2012tensor, wetzstein2011layered3d, lee2016additivelf, wetzstein2012compressivelf, maimone2013focus3d, huang2015stereoscope}. These systems typically employ either (1) a stack of attenuating layers that modulate a backlight panel, known as \textit{multiplicative} displays, where the final image corresponds to the product of all layer transmittances, or (2) a stack of diffusive or emissive layers, known as \textit{additive} displays, where the resulting image is formed by summing the contributions from each layer. The display patterns for each layer are determined by solving a nonnegative tensor factorization problem such that the combined emission or transmission of all layers reproduces a target light field. As the viewer moves laterally, the layered stack generates different angular views, thereby reconstructing a full 4D light field.

In computer graphics and, more recently, neural rendering, similar \textit{layered 3D representations} have been proposed to achieve efficient light field synthesis. Notably, the \textit{multiplane images (MPIs)} representation~\cite{zhou2018mpi} models a scene as a set of RGBA images positioned at discrete depth planes. MPIs are compact, differentiable, and renderable in real time using simple planar homographies and alpha compositing. Their fronto-parallel structure enables high-fidelity reconstruction of forward-facing light fields with moderate parallax using only a handful of depth layers~\cite{flynn2019deepviewmpi, srinivasan2019pushingmpi, pu2023sinmpi, khan2023tiledmpi, han2022adaptivempi, tucker2020singlempi, zhao2022generativempi, luvizon2021adaptivempi}. These characteristics are particularly advantageous for holographic near-eye displays, which primarily require accurate reconstruction of front-facing views in a small eyebox rather than full 360° viewing coverage, while also demanding real-time CGH computation.

Motivated by these insights, we propose CGH algorithms that extend layered representations such as MPIs into the wave-optics domain. Our method converts such layered 3D representations into complex holograms that reconstruct full 4D light fields. Layered formats are inherently well-suited for efficient hologram synthesis: they require only a limited number of depth planes (typically fewer than 100) to be propagated, in contrast to the thousands or millions of primitives processed in Gaussian- or polygon-based holography. This structural sparsity leads to orders-of-magnitude acceleration in computation. We demonstrate that optimized layered representations enable hologram synthesis with up to $250,000\times$ speedup compared to state-of-the-art primitive-based CGH algorithm, Random-phase Wave Splatting (RPWS) \cite{chao2025rpws}, while maintaining excellent light field reconstruction quality.

\paragraph{\bf{Complex-valued Holographic Radiance Fields.}} In a concurrent work, \citet{zhan2025complexradiance} proposed \textit{Complex-valued Radiance Fields} that optimizes a scene-specific complex representation in which Gaussian primitives carry complex features and are composited along a set of discrete planes during rendering. This yields a layered, plane-wise accumulation procedure reminiscent of multiplane formulations, but the underlying parameterization is a completely new, learned complex scene model. In contrast, our approach operates directly on off-the-shelf layered inputs (e.g., existing multiplane images) and targets hologram synthesis with full $2\pi$ random phase and time multiplexing for \textit{both} focal stack and light-field reconstruction and wide-eyebox viewing.

%% file: main_paper/sections/3_method.tex
\begin{figure}
    \centering
    \includegraphics[width=\linewidth]{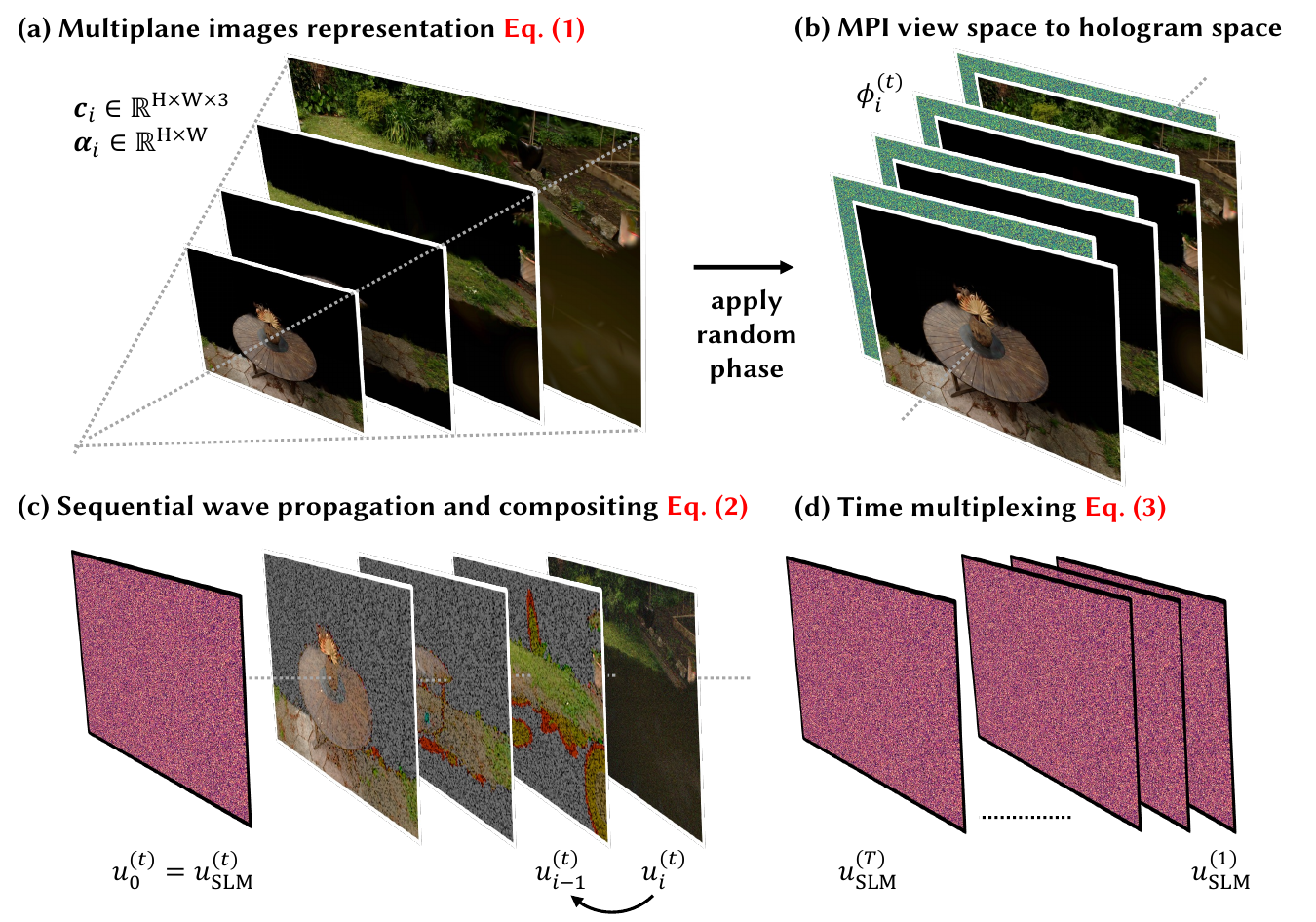}
    \caption{\textbf{The MPI CGH pipeline.} Our MPI-based CGH pipeline consists of 4 steps. We start with a set of \textbf{(a)} camera frustum-aligned multiplane images defined by Eq. \ref{eq:mpi_representation}, and \textbf{(b)} transform them into complex wavefronts aligned with the SLM plane by applying a random phase map to each RGBA image slice. The complex MPI wavefronts are \textbf{(c)} sequentially propagated and composited at each MPI depth plane using our novel wave compositing procedure defined by Eq. \ref{eq:mpi_random_tm} to get a single-frame MPI hologram at the SLM plane. Finally, multiple MPI holograms with different random phase patterns are \textbf{(d)} time multiplexed as shown in Eq. \ref{eq:tm}. }
    \label{fig:mpi_pipeline}
\end{figure}

Our approach converts existing multiplane images to \textit{random-phase} holograms that accurately reconstruct 3D focal stacks and 4D light fields. We first describe the MPI image representation in Section \ref{subsec:mpi_representation}, and present our MPI-based CGH rendering pipeline in \ref{subsec:mpi_cgh}. The overall MPI-based CGH pipeline is outlined in Fig. \ref{fig:mpi_pipeline}.

\subsection{Multiplane Image Representation}
\label{subsec:mpi_representation}

Given a reference camera viewpoint (camera extrinsics defined by the camera translation vector $\mathbf{t} \in \mathbb{R}^3$ and rotation matrix $\mathbf{R}\in \mathbb{R}^{3\times3}$) and the camera intrinsics $\mathbf{K} \in \mathbb{R}^{3\times3}$, the \textit{multiplane images (MPIs)} with respect to the given camera viewpoint are a set of $L$ of RGBA images $(\mathbf{c}_i, \boldsymbol{\alpha}_i)$ defined at different depths $d_i \in \mathbb{R}, 1\leq i \leq L$ from the camera:
\begin{align}
    \{(\mathbf{c}_i, \boldsymbol{\alpha}_i) \mid \mathbf{c}_i \in \mathbb{R}^{H \times W \times 3}, \ \boldsymbol{\alpha}_i \in \mathbb{R}^{H \times W}, \ i = 1, \dots, L\}.
    \label{eq:mpi_representation}
\end{align}
as shown in Fig. \ref{fig:mpi_pipeline}. 

It has been shown in prior work that optimized MPIs can reconstruct local, front-facing light fields around the reference camera viewpoint exceptionally well \cite{zhou2018mpi}. To render a novel view from a new target camera viewpoint $\mathbf{t}', \mathbf{R}'$ given the MPIs with respect to a reference camera viewpoint $\mathbf{t}, \mathbf{R}$, each MPI layer is warped to the target viewpoint using simple planar homography. The warped MPI layers $(\mathbf{c}'_i, \boldsymbol{\alpha}'_i), 1\leq i \leq L$ are then alpha-composited from back-to-front using the standard \textbf{over} operation \cite{porter1984over}. 

\begin{figure*}[!th]
    \centering
    \includegraphics[width=\textwidth]{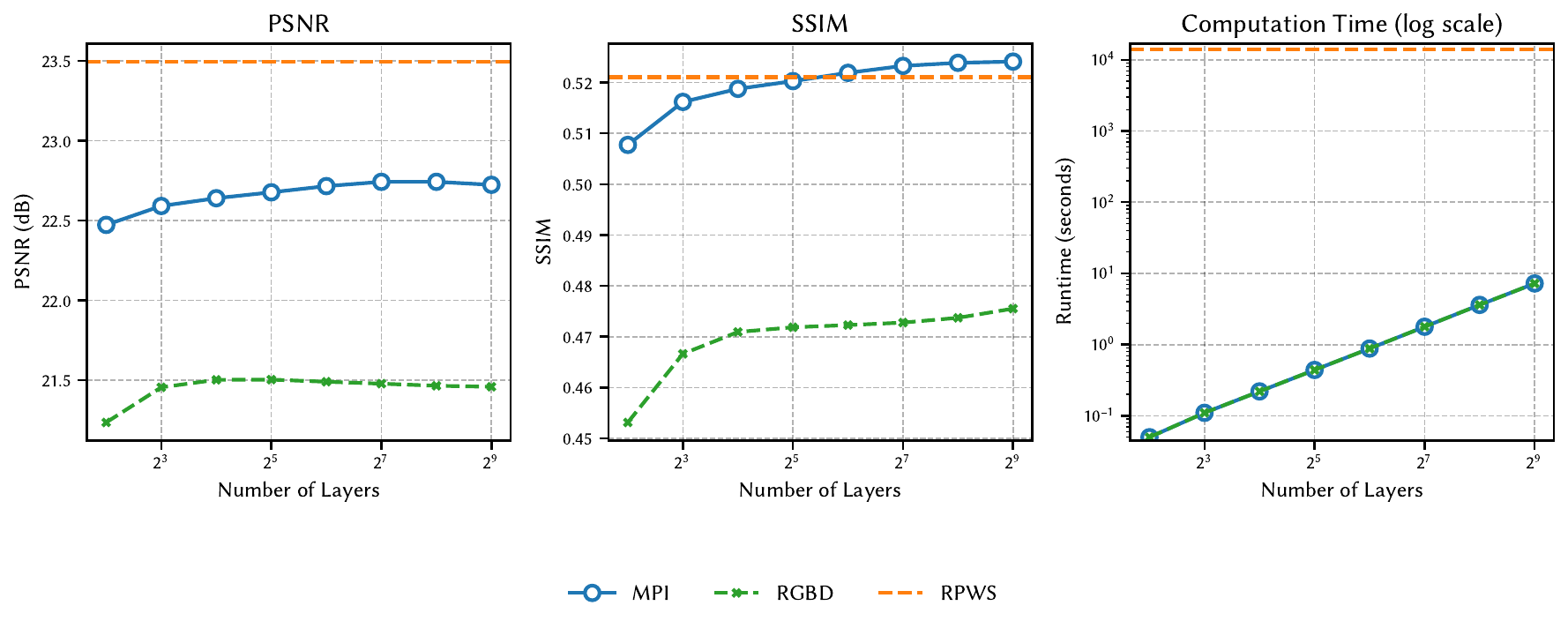}
    \caption{\textbf{Runtime and image quality analysis of different CGH algorithms.} We compare runtime and light field reconstruction quality across CGH algorithms with varying numbers of depth planes for RGBD- and MPI-based methods. We report average metrics on holograms generated from selected scenes of the Mip-NeRF 360 \cite{barron2022mipnerf360} dataset. RPWS achieves the highest image fidelity but is computationally expensive, being more than $2000\times$ slower than layer-based approaches (MPI and RGBD), whose runtime scales approximately linearly with the number of depth layers. Crucially, the computational cost is dictated solely by the layer count, being nearly identical for both MPI and RGBD representations, represented by the overlaid blue and green lines in the runtime plot. RGBD holograms exhibit low image quality that improves with more depth planes yet remains inferior to MPI holograms. In contrast, MPI-based CGH attains a favorable tradeoff, rendering up to $250,000\times$ faster while matching RPWS in SSIM and surpassing RGBD in overall visual quality.}
    \label{fig:metrics_plot}
\end{figure*}

\subsection{Wave Optics Rendering of MPI Images}

Given a set of multiplane images aligned with the center camera viewpoint and parallel to the \textit{SLM plane} in the holographic display, we transform MPIs to holograms by modulating RGBA images with random phase masks and sequentially propagate and blend them from back to front to get the final composited complex wavefront on the SLM plane. 

We formulate the rendering process recursively in Eq. \ref{eq:mpi_random_tm}, where $u_i(\coordinate)$ represents the accumulated complex field emerging from layer $i$ after integrating all background layers from $N$. Given the opacity channel $\boldsymbol{\alpha}_i(\coordinate)$ of the current slice, we propagate the resulting wavefront across the inter-layer gap $\Delta z = z_{i - 1} - z_{i}$ to the subsequent plane $i-1$ using the angular spectrum method \cite{goodman2005introduction} and composite them using a recently proposed wave optics alpha blending formulation for random-phase wavefronts \cite{chao2025rpws}:

\begin{align}
    \hspace*{-0.5em}
    u_{i - 1}^{(t)} (\coordinate) = \mathcal{P} \Big( \sqrt{1 - \boldsymbol{\alpha}_i(\coordinate)} \; u_i(\coordinate) +  \sqrt{c_i} \sqrt{\boldsymbol{\alpha}_i(\coordinate)} \; e^{i \phi_i^{(t)}(\coordinate)} ; \Delta z\Big)
    \label{eq:mpi_random_tm}
\end{align}
We define $t \in \{1, \dots, T\}$ as the frame index of a time-multiplexed MPI hologram with $T$ frames. For each layer $i$ in frame $t$, a unique random phase profile $\phi_i^{(t)}$ is applied. The recursion terminates at the display plane ($z_0=0$), where the target complex field is given by $u_\text{SLM}^{(t)}(\mathbf{x}) = u_0^{(t)}(\mathbf{x})$.

We compute the time-averaged intensity $I(\mathbf{x})$ of the final time-multiplexed MPI hologram by accumulating the intensity of the reconstructed field under a generalized operator $\mathcal{O}(\cdot; \cdot)$:
\begin{align}
    I(\mathbf{x})  = \frac{1}{T} \sum_{t=1}^{T} \left| \mathcal{O} \left( u_\text{SLM}^{(t)}(\mathbf{x}); \cdot \right) \right|^2 ,
    \label{eq:tm}
\end{align}
This formulation is agnostic to the operator $\mathcal{O}(\cdot; \cdot)$, which may represent standard free-space propagation for 2D imaging or 3D focal stacks, the Short-time Fourier Transform (STFT) for 4D light field reconstruction~\cite{choi2022time}, or even pupil masking functions for synthesizing specific angular views~\cite{schiffers2023stochastic, shi2024ergonomic}, which we show in Section 4.

With this MPI-based CGH rendering pipeline, we transform MPIs, which are traditionally limited to ray-optics-based rendering of viewpoint-dependent 2D images, into complex-valued holograms that accurately reconstruct 3D focal stacks and 4D light fields under a full wave-optics image formation model, as demonstrated in the following section. 

\label{subsec:mpi_cgh}

%% file: main_paper/sections/4_results.tex
In this section, we describe implementation details including the data preparation pipeline and hardware configurations in Section \ref{subsec:impl} and show simulated and experimentally captured results in Section \ref{subsec:results}.

\subsection{Implementation Details}
\label{subsec:impl}

\subsubsection{Evaluation Datasets} We demonstrate the effectiveness of our method on a wide collection of scenes from the NeRF Blender synthetic dataset \cite{mildenhall2020nerf} and the Mip-NeRF 360 \cite{barron2022mipnerf360} dataset. 

\subsubsection{Baseline 3D Scene Representations} We compare MPI-based CGH with conventional layer-based CGH \cite{peng2020neural, shi2021towards, shi2022end} using RGBD images, and the state-of-the-art primitives-based random-phase CGH method, Random-phase Wave Splatting (RPWS) with 2D Gaussians \cite{chao2025rpws}.

For the RPWS baseline, we use the \texttt{gsplat} software library to optimize 2DGS representations for Gaussian-based CGH. As described in RPWS \cite{chao2025rpws}, we optimize Gaussians to reconstruct the square of the target image, and extract the square root of the optimized Gaussian colors of CGH calculation.

\begin{figure*}[t!]
    \centering
    \includegraphics[width=\textwidth]{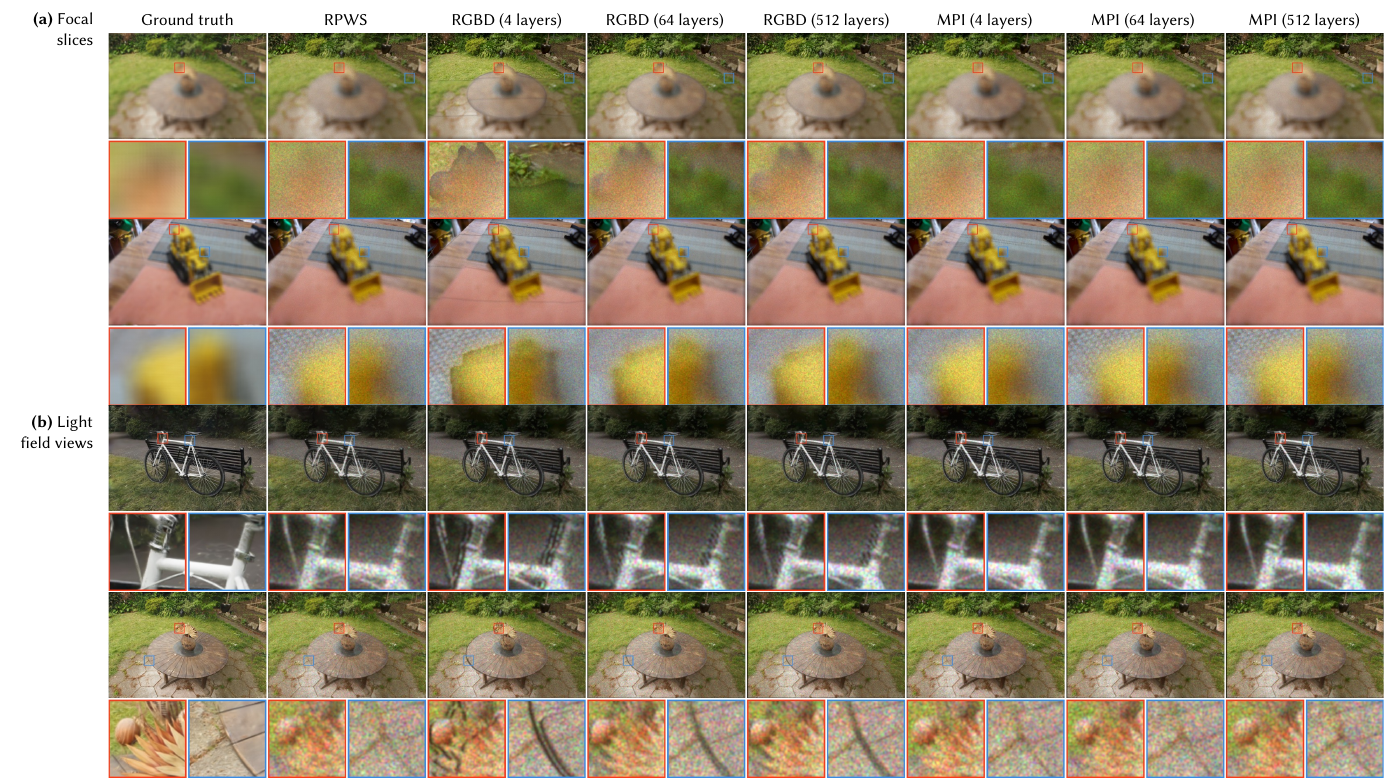}
    \caption{\textbf{Defocus and occlusion behaviours of different CGH algorithms at depth transitions.} We show zoom-in crops of 3D focal stacks and 4D light fields reconstructed using different CGH algorithms. RPWS, a state-of-the-art primitives-based CGH method, produces the most natural defocus blur and accurate occlusion behavior. In contrast, RGBD holograms exhibit light leakage and halo artifacts in defocus regions, as well as dark occlusion boundaries in light field views. Increasing the number of RGBD depth layers does not alleviate these issues. Our MPI holograms, on the other hand, reconstruct natural defocus blur and accurate occlusion effects using as few as four depth layers. }
    \label{fig:artifacts_comparison}
\end{figure*}

There are many ways to generate an MPI representation of a 3D scene. Prior methods include training a feed-forward neural network to predict the MPIs of a 3D scene given a stereo image pair \cite{zhou2018mpi} or from a single image \cite{pu2023sinmpi, khan2023tiledmpi, tucker2020singlempi, han2022adaptivempi}.

Since our goal is to demonstrate the potential of MPIs for CGH, we do not focus on how to acquire the MPI representation itself as there has already been extensive research in such area and that \textit{our method is agnostic of the MPI generation process}. Instead, we simply extract MPIs from a set of pre-optimized Gaussian splats by binding the Gaussians to predetermined depth planes. Specifically, given a set of depth-sorted Gaussian splats, we define the color image $c_{d_k}$ and alpha image $\alpha_{d_k}$ at depth plane $d_k, 1\leq k\leq D$ to be 
\begin{align}
    c_{d_k} & = \frac{\sum_{i=s_k}^{s_k+N_k}\alpha_i c_i\prod_{j<i}(1-\alpha_j)}{1 - \prod_{i=s_k}^{N_k}(1-\alpha_i)}, \;
    \alpha_{d_k} = 1 - \prod_{i=s_k}^{N_k}(1-\alpha_i)
\end{align}
We partition the depth planes such that they are uniformly distributed within $(z_\text{min}, z_\text{max})$, which are the nearest and farthest depth planes of the 3D hologram volume, respectively.

\begin{figure}[tb]
    \centering
    \includegraphics[width=\linewidth]{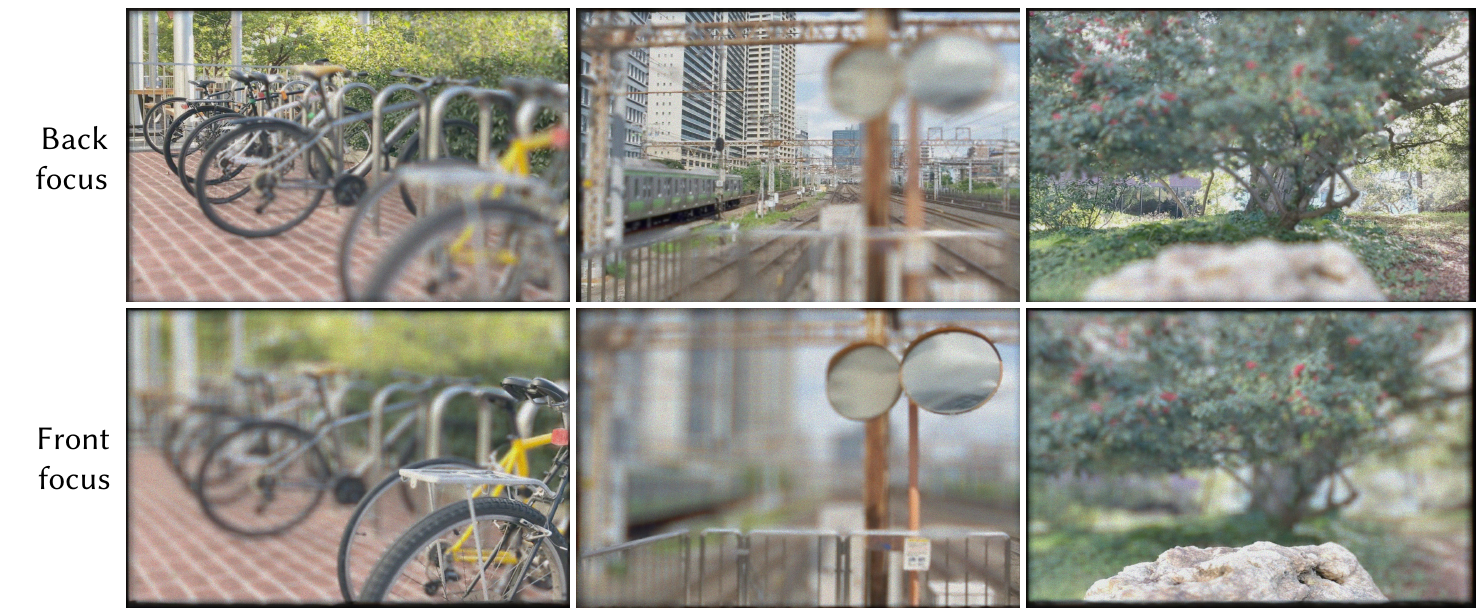}
    \caption{\textbf{MPI-based CGH works with MPIs produced by off-the-shelf neural networks.} We use MPIs generated by Stereo Magnification~\cite{zhou2018mpi}, a representative neural-network–based MPI predictor, for hologram synthesis. Our MPI-based CGH algorithm is \textit{agnostic to the MPI generation process} and reconstructs accurate 3D focal stacks.
}
    \label{fig:stereo_mag}
\end{figure}

Most importantly, this data generation pipeline allows us to flexibly control the number of layers in the MPI representation, which is not supported for current data-driven MPI generation networks as they all generate MPIs with a fixed number of layers. This flexibility allows for extensive ablation studies and benchmarking experiments in Section 4 to help investigate the effect of the number of MPI layers on image quality and CGH runtime. We addtionally demonstrate that MPI-based CGH is independent of how MPIs are acquired in Fig. \ref{fig:stereo_mag}, where we show qualitative results of MPI holograms generated using a representative neural-network-based MPI predictor, Stereo Magnification \cite{zhou2018mpi}. 
 
To generate RGBD images of a given view, we use DepthAnything \cite{depth_anything_v2} to predict its depth and partition the scene into different depth bins to generate corresponding depth masks and masked RGB images. We use the same uniform depth partitioning scheme as MPIs.

\subsubsection{Algorithm Implementations. } We implement MPI-based CGH and all baseline methods in PyTorch. Please refer to the supplementary materials for pseudocode and implementation details.


\begin{figure*}[t!]
    \centering
    \includegraphics[width=\textwidth]{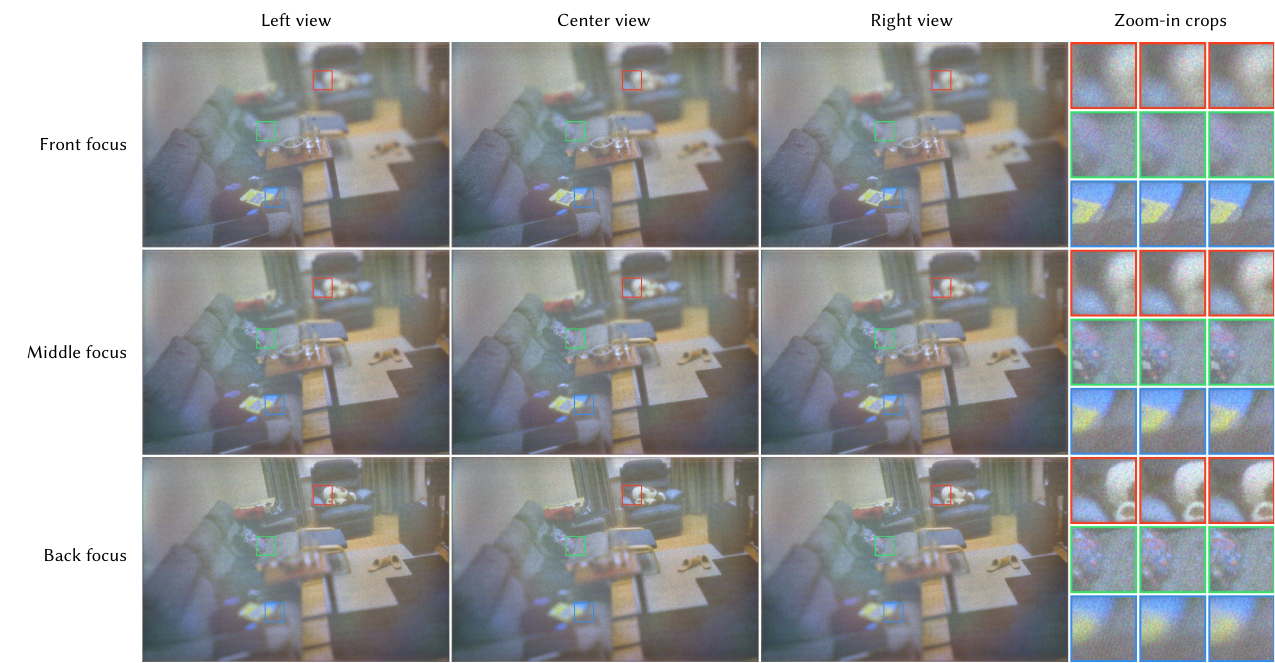}
    \caption{\textbf{Experimentally captured light field parallax at different focused depths.} We capture light field parallax \textit{at different depth planes} by translating both the pupil horizontally at the Fourier plane and the camera along the optical axis. As expected, the parallax shift of objects increases as they move farther away from the focused depth, as seen from the red zoom-in crops for the front-focus results, and the blue zoom-in crops from the back-focus results. The objects that are at the focused depth, i.e. light field plane, don't move. These results collectively demonstrate that our MPI holograms reconstruct accurate light fields that respect standard ray optics geometry.}
    \label{fig:lf_different_depths}
\end{figure*}

\subsubsection{Experimental Hardware Setup.} MPI-based CGH generates complex-valued holograms. We perform SGD \cite{peng2020neural} with a complex field supervision loss \cite{chen2021complex} to synthesize phase-only holograms that accurately reconstruct the target complex wavefront at a fixed propagation distance (4cm) from the phase-only SLM. We use a 1080p HOLOEYE Pluto-2.1 phase-only SLM with $8 \mu m$ pixel pitch and a FISBA READYBeam Ind2 fiber-coupled laser module as the illumination source. We use a FLIR GS3-U3-123S6C-C color camera placed on a z-axis translation stage (Thorlabs MTS25-Z8 50mm) to capture focal slices at different distances from the SLM. For horizontal parallax and light field capture, we place an adjustable pupil on a motorized horizontal translation stage (Newport Optics CONEX-AGP) at the Fourier plane. We refer readers to the supplementary materials for more details on the experimental setup.

\subsection{Simulation and Experimental Results}
\label{subsec:results}
\subsubsection{Baseline Comparisons with Simulation Results. } We compare our method (MPI) with prior layer-based RGBD CGH and a state-of-the-art primitives-based CGH algorithm, Random-phase Wave Splatting (RPWS) with 2D Gaussians \cite{chao2025rpws}. For RGBD-based CGH and our MPI-based CGH method, we also ablate the number of depth layers. We use 24 frames time multiplexing ($T=24$) for all baselines.

In Fig. \ref{fig:artifacts_comparison} and \ref{fig:results_sim}, we see that RPWS achieves the best quantitative 3D focal stack and 4D light field reconstruction quality at the cost of slow runtime. On the other hand, RGBD-based CGH achieves fast runtime yet results in poor reconstruction quality. Our MPI-based CGH pipeline gets the best of both worlds, achieving \textit{both} fast runtime and high image quality. RGBD holograms reconstruct somewhat plausible focal stacks and light fields, yet severe halo, leakage, and dark border artifacts, often appearing as “depth-tearing” near depth discontinuities, remain prominent. RPWS holograms achieve high image quality through dense 2D Gaussian representations with millions of primitives, but incur extremely slow runtimes. In contrast, our MPI holograms reconstruct both focal stacks and light fields with perceptually similar quality to RPWS while simultaneously achieving fast runtime.

Additionally, Fig. \ref{fig:metrics_plot} presents runtime and image quality comparisons across various CGH baselines. Both RGBD-based CGH and MPI-based CGH reconstruction quality improve with an increasing number of depth layers, but their runtime scales approximately linearly with layer count. While this reflects an inherent trade-off between quantitative image quality and computational cost, we observe that perceptual quality saturates beyond roughly 30 layers, suggesting that MPI-based CGH offers a favorable balance for interactive display applications.

\subsubsection{Experimentally Captured Focal Stack Results. } In Fig. \ref{fig:results_captured}, we show experimentally captured 3D focal stacks of holograms synthesized using different CGH algorithms. The captured results closely follow the trend observed in our simulations. MPI holograms reconstruct focal stacks with accuracy comparable to RPWS, whereas RGBD holograms struggle to reproduce realistic defocus blur even with additional depth layers.

\subsubsection{Experimentally Captured Light Field Parallax Results. } In Fig. \ref{fig:results_captured}, we show experimentally captured horizontal parallax-only light fields of holograms synthesized using different CGH algorithms when the camera is focused at the hologram plane, or middle depth plane. We also capture the parallax effects at different focal planes (rear, middle, and front planes), which is shown in Fig. \ref{fig:lf_different_depths} and the supplemental materials. The captured light field results also closely follow the trend observed in our simulations, where we see clear parallax shifts in the zoom-in crops.

%% file: main_paper/sections/5_discussion.tex
\noindentparagraph{\bf{Limitations and Future Work.}}
Currently, our method achieves up to $250,000\times$ speedup over state-of-the-art primitives-based CGH methods, but is still not real-time. Combining MPI-based CGH with recent real-time random phase hologram synthesis networks \cite{artifact2025chu} could be an interesting direction. The use of random phase leads to lower contrast in the reconstructed images. A promising solution to this issue are learning-based calibration methods \cite{peng2020neural, choi2022time, jang2024waveguide, choi2025sah}. Please refer to the supplemental materials for more discussions on contrast loss in random phase holograms.

\noindentparagraph{\bf{Conclusion.}} We propose MPI-based CGH, a novel CGH pipeline based on multiplane images (MPIs) that achieves high focal stack and light field reconstruction quality on par with state-of-the-art primitives-based CGH methods while greatly speeding up CGH runtime by up to $250,000\times$. This significant advance in rendering efficiency paves the way for future real-time computer-generated holography for interactive AR and VR experiences.